\documentclass[aps,pra,floatfix,twocolumn,showpacs]{revtex4}
\usepackage{epsfig}
\usepackage{graphicx}
\usepackage{dcolumn}
\begin{document}
\title{Fine Structure Anomalies and Search for Variation of the Fine
Structure Constant in Laboratory Experiments}
\author{V. A. Dzuba$^1$}
\email{V.Dzuba@unsw.edu.au}
\author{V. V. Flambaum$^1$$,$$^{2}$}
\email{V.Flambaum@unsw.edu.au}
\affiliation{$^1$School of Physics, University of New South Wales, Sydney 2052,
Australia\\
    $^2$ Physics Division, Argonne National Laboratory, Argonne, Illinois 60439-4843, USA }

\date{\today}

\begin{abstract}

Configuration interaction in many-electron atoms may cause anomalies in 
the fine structure which make the intervals
 small and very sensitive
to  variation of the fine structure constant. Repeated precision  
measurements of these intervals over long period of time can put strong 
constrain on possible time variation of the fine structure constant.
We consider the $5p^4 \ ^3$P$_{2,1,0}$ fine structure multiplet in the 
ground state of neutral tellurium as an example. 
Here the effect of change of the fine structure constant is
enhanced about one hundred times in the relative change of the small
energy interval between the $^3$P$_1$ and  $^3$P$_0$ states.

\end{abstract}


\maketitle

\section{introduction}

The possibility of fundamental constants to vary is predicted by theories
unifying gravity with other interactions 
(see, e.g. review by Uzan~\cite{uzan}). There is an evidence found in
quasar absorption spectra that the fine structure constant $\alpha$ 
($\alpha = e^2/\hbar c$) might be smaller in early cosmological 
epoch~\cite{1,2,3,4}. However, similar analysis applied by other authors to 
different sets of data does not support this claim \cite{Srianand,Quast}. 
Recent progress in improving atomic clocks and developing optical frequency 
standards (see, e.g. \cite{clocks}) made it possible to put strong 
constrains on possible time variation of fundamental constants in laboratory
measurements. In particular, constrains obtained for the fine structure 
constant  $\alpha$ need only about one order of magnitude improvement 
to see whether they are consistent with the quasar absorption spectra
data if the same rate of change is assumed over cosmological time.
These constrains were obtained with the full use of the technique
developed for highly accurate optical standards. However, the
choice of atomic transitions was not optimal for searching of variation 
of the fine structure constant. Frequencies used so far change at about
the same rate as $\alpha$ if $\alpha$ changes, i.e. there is no enhancement.

An alternative approach was suggested in Refs.\cite{flambaum1,flambaum2}.
It was proposed to measure small frequency between two close states with 
different dependence on $\alpha$.
 Here small change of $\alpha$ may lead to
orders of magnitude larger 
relative
 change in frequency. The enhancement
 factor $k$
can be expressed as
 $k = 2 q_{12}/\omega$, where $ q_{12}$ is the 
difference in relativistic energy shifts of two levels and $\omega$ is
transition frequency. A good candidate for such experiments is dysprosium
atom~\cite{flambaum1,flambaum2,budker}. It has two almost degenerate states 
of opposite parity and corresponding enhancement factor is of order
of $10^{8}$~\cite{flambaum2}. The measurements for dysprosium are currently
underway at Berkeley~\cite{budker}.

While it is
 hard to find anything better than dysprosium in terms of 
enhancement factor, it has its disadvantages too. The levels involved
are not so narrow. While one of the states lives long  enough
to cause no problem the other level does not.
 One has to work inside the line width
to have the desirable accuracy. This
 puts
 certain limitations on achievable
constrains on $\alpha$-variation. Therefore, it would be important to find
something which combines the advantages of having metastable states and
strong enhancement. A number of such transitions were considered in
recent work~\cite{saveliy}. It has been suggested to look at close
metastable states of different configurations. Energies of different
configurations change at different rates when $\alpha$ is changing. 
This, together with small energy intervals between the states ensures
strong enhancement~\cite{saveliy}.

In present work we consider slightly different possibility. We suggest
to study anomalously small
fine structure intervals in ground configurations of many-electron atoms. 
The use of the ground state fine structure multiplet ensures that states 
are metastable. They can only decay via $M1$-transitions which are very 
much suppressed due to small value of transition frequency. On the other 
hand, configuration
interaction in many-electron atoms can  
reduce fine structure intervals and lead to strong sensitivity  to
the  change in the fine structure constant. In next sections we consider
 in detail fine structure of the ground state of tellurium and discuss
 some other possibilities.

\section{Theory and results for tellurium}

To study sensitivity of atomic frequencies to variation of the fine
structure constant $\alpha$ it is convenient to present them in the 
vicinity of the physical value of $\alpha$ ($\alpha = \alpha_0 = 1/137.036$) 
in the form
\begin{equation}
  \epsilon = \epsilon_0 + qx,
\label{omegaq}
\end{equation}
where $x=(\alpha/\alpha_0)^2-1$ and coefficient $q$ defines the sensitivity
of the frequency to variation of $\alpha$. In general, its value can be 
found from atomic calculations. If $\alpha$ changes
the relative
 frequency changes at
the rate
\begin{equation}
  \frac{\Delta \omega}{\omega} = \frac{2q_{12}}{\omega_0} 
  \frac{\Delta \alpha}{\alpha_0} \equiv k \frac{\Delta \alpha}{\alpha_0},
\label{kfact}
\end{equation}
where $k=2q_{12}/\omega_0$ is enhancement factor. To search for variation of 
the fine structure constant one needs to compare atomic frequencies
with different enhancement factors over long period of time. The larger
this difference the more sensitive the experiment to variation of $\alpha$.
Note that $k$ can have negative value which means that changes of $\alpha$
and frequency go in opposite directions: frequency decreases when $\alpha$
increases and vise versa.

For ``normal'' fine structure intervals which are proportional to 
$(Z\alpha)^2$ formula (\ref{omegaq}) is valid for all values of $\alpha$,
$0 < \alpha < \alpha_0$. Therefore,
 $q_{12}=\omega_0$ and $k=2$.
In other words, the factor $k$
 is the same for all ``normal'' fine structure intervals
and comparison between them cannot reveal any drift of $\alpha$.
These fine structure intervals can still be used in search for variation
of
 the fundamental constants
 if they are compared to hyperfine structure or to frequencies
of suitable optical transitions.

Situation changes dramatically if fine structure multiplet is strongly
perturbed by configuration interaction with neighboring states.
We consider neutral tellurium atom in its ground state as an example.

Experimental and theoretical energies of the ground state $5p^4$
configuration of tellurium are presented in Table~\ref{tab1}.
There are strong anomalies in the fine structure multiplet $^3$P$_{2,1,0}$.
The $^3$P$_2$ -  $^3$P$_1$ and  $^3$P$_1$ -  $^3$P$_0$ intervals have
opposite signs and differ in value more than hundred times. This is
because of configuration interaction between the $^3$P$_2$ and $^1$D$_2$
states and between the  $^3$P$_0$ and  $^1$S$_0$ states while the $^3$P$_1$
state has no close neighbors to mix with. Configuration interaction leads
to accidental almost exact cancellation between spin-orbit and Coulomb
terms in the energy interval between the  $^3$P$_1$ and  $^3$P$_0$ states.
Since spin-orbit
interaction
 is much more sensitive to the change of $\alpha$ than
the Coulomb term, it is natural to expect that the  $^3$P$_1$ - $^3$P$_0$
energy interval is very sensitive to the variation of $\alpha$.

To check how strong is sensitivity of the fine structure intervals
of Te~I we perform model configuration interaction (CI) calculations
for the $5p^4$ configuration of the atom. First, we perform Hartree-Fock
calculations 
for open shells  to find the $5p_{1/2}$ and $5p_{3/2}$
states of neutral tellurium. Then we use the CI technique to construct
four-electron states of the $5p^4$ configuration and to calculate their
energies
 (in fact, in this approximation CI technique is reduced
to diagonalization of the interaction Hamiltonian describing
direct mixing of different $5p^4$ states).
 It turns out that some extra fitting is needed to have good 
agreement with experiment. Namely, we reduce the value of the 
$F_2(5p_{3/2},5p_{3/2})$ Coulomb integral by 25\%. This reduction 
simulates the effect of screening of Coulomb interaction between 
valence electrons by core electrons. The results for energies are 
presented in Table~\ref{tab1}. One can see that in spite of simple 
approximation used in calculations the agreement with experiment
is very good.

Coefficients $q$ (see Eq.(\ref{omegaq})) are found by varying
$\alpha$ in computer codes:
\[   q=\frac{\epsilon(x=+0.1)-\epsilon(x=-0.1)}{0.2}. \]
The results for $q$ are also presented in Table~\ref{tab1}. Since
we have good agreement with experiment for the energies it is natural
to assume that the accuracy for the $q$-coefficients is also good.

\begin{table}
\caption{\label{tab1}Experimental and theoretical energies and 
$q$-coefficients (cm$^{-1}$) for the $5p^4$ ground-state configuration 
of Te~I.}
\begin{ruledtabular}
\begin{tabular}{ccrrr}
State & $J$ & \multicolumn{1}{c}{$E$(exp)} & \multicolumn{1}{c}{$E$(theor)}
& \multicolumn{1}{c}{$q$} \\
\hline
$5p^4 \ \ ^3P$ & 2 &     0 &     0 &    0 \\
               & 1 &  4751 &  4770 & 5927 \\
               & 0 &  4707 &  4713 & 3594 \\
$5p^4 \ \ ^1D$ & 2 & 10559 & 10745 & 5207 \\
$5p^4 \ \ ^1S$ & 0 & 23199 & 22845 & 8571 \\
\end{tabular}
\end{ruledtabular}
\end{table}

Table~\ref{tab2} presents frequencies and enhancement factors for transitions
between all states of the $^3$P$_{0,1,2}$ fine structure multiplet. The 
enhancement for the $^3$P$_1 \ - \ ^3$P$_0$ is more than one hundred due
to anomalously small frequency of the transition. 
Ratio of
 this small frequency to almost any other atomic frequency
is extremely sensitive to variation of $\alpha$. However, other transitions
from Table~\ref{tab2} can also be used.

\begin{table}
\caption{\label{tab2}Frequencies (in cm$^{-1}$) and enhancement factors 
($k=2\Delta q/\omega$)
for transitions within the $^3$P$_{0,1,2}$ fine structure multiplet of Te~I}. 
\begin{ruledtabular}
\begin{tabular}{ccrr}
Transition & Type & \multicolumn{1}{c}{$\omega$} & \multicolumn{1}{c}{$k$} \\
\hline
$^3$P$_1 \ - \ ^3$P$_2$ & M1 &  4751 &   2.2 \\
$^3$P$_0 \ - \ ^3$P$_2$ & E2 &  4707 &   1.5 \\
$^3$P$_1 \ - \ ^3$P$_0$ & M1 &    44 &   106 \\
\end{tabular}
\end{ruledtabular}
\end{table}

\begin{figure}
\centering
\epsfig{figure=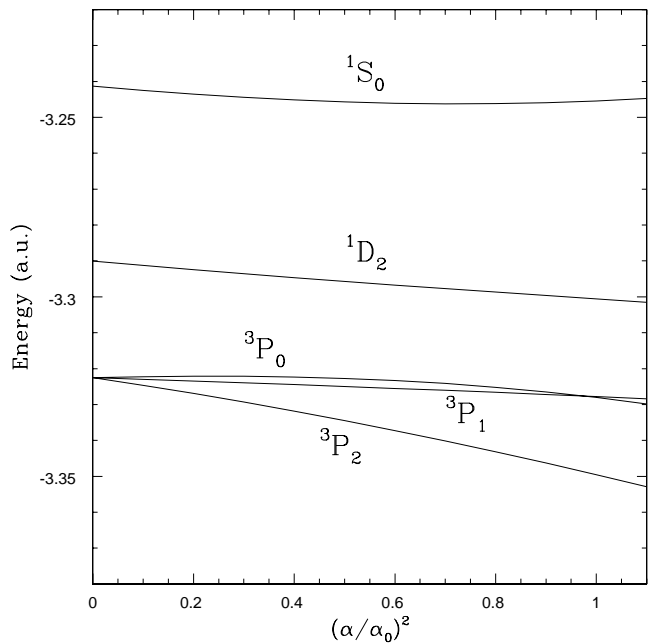,scale=0.45}
\caption{Energy levels of lower states of Te~I as functions of the fine
structure constant.}
\label{fig}
\end{figure}

Because of very exotic behavior of the fine structure intervals of Te~I 
it is very interesting to see what happens to them when $\alpha$ varies
from zero (non-relativistic limit) to its physical value $\alpha_0$.
We have performed such calculations and results for five low states of the
$5p^4$ ground-state configuration of Te~I are presented on 
Fig.~\ref{fig}. Valence energies (energy to remove
all four valence electrons from atom) are shown as functions of
$(\alpha/\alpha_0)^2$. All three energies of the $^3$P$_{0,1,2}$
multiplet have the same value at $\alpha=0$ and fine structure intervals
are proportional to $\alpha^2$ at small $\alpha$. At larger values 
of $\alpha$ interaction
with the $^1$D$_2$ and $^1$S$_0$ leads to significant perturbation of the 
$^3$P multiplet. In particular, repulsion between the $^1$S$_0$ and
$^3$P$_0$ levels
 causes
 the latter
 to cross with the $^3$P$_1$ level
in the vicinity of the physical value of $\alpha$. This explains
anomalously small energy interval between the two states.

Let us mention in the end of this section that both upper states of the
$^3$P multiplet are metastable. The $^3$P$_1$ state decays mostly by M1
transition to the ground state. Its lifetime is about 1 sec. 
The $^3$P$_0$ state decays via E2 transition to the ground state and 
corresponding lifetime (if no other factors are involved) is about 
$4 \times 10^3$ sec.

\section{Polonium and cerium}

There are many other examples of the anomalous fine structure in the 
ground and low excited states of many-electron atoms which involve 
metastable states and large enhancement and therefore suitable for
search of variation of the fine structure constant. The actual choice
between them would involve consideration of many other factors which 
are important for experimentalists but not discussed in present paper.
Below we discuss two more examples just to have broader picture.

Brief look at the spectra of elements presented in Moore's tables~\cite{moore}
reveals that practically all elements with the $np^4$ configuration in the
ground state have anomalies in the fine structure (though none of them has 
so small energy interval as Te~I). The most interesting case after Te~I
is probably polonium (Po~I). It has the largest nuclear charge $Z$ ($Z=84$) 
among the $np^4$ elements which mean strong relativistic effects and 
large $q$ and $k$ coefficients (see formulas (\ref{omegaq}) and (\ref{kfact})).
Experimental and theoretical data similar to those considered for Te~I
are
 presented in Tables \ref{tabpo} and \ref{kpo}. We see that enhancement
factors $k$ are large and different for different transitions. This is
exactly what is needed for the search of variation of the fine structure 
constant.

\begin{table}
\caption{\label{tabpo}Experimental and theoretical energies and 
$q$-coefficients (cm$^{-1}$) for the $6p^4$ ground-state configuration 
of Po~I.}
\begin{ruledtabular}
\begin{tabular}{ccrrr}
State & $J$ & \multicolumn{1}{c}{$E$(exp)} & \multicolumn{1}{c}{$E$(theor)}
& \multicolumn{1}{c}{$q$} \\
\hline
$6p^4 \ \ ^3P$ & 2 &     0 &     0 &     0 \\
               & 1 & 16831 & 17248 & 26720 \\
               & 0 &  7514 &  7397 &  1910 \\
$6p^4 \ \ ^1D$ & 2 & 21679 & 22189 & 25850 \\
$6p^4 \ \ ^1S$ & 0 & 42718 & 45101 & 54130 \\
\end{tabular}
\end{ruledtabular}
\end{table}

\begin{table}
\caption{\label{kpo}Frequencies (in cm$^{-1}$) and enhancement factors 
($k=2\Delta q/\omega$)
for transitions within the $^3$P$_{0,1,2}$ fine structure multiplet of Po~I.} 
\begin{ruledtabular}
\begin{tabular}{ccrr}
Transition & Type & \multicolumn{1}{c}{$\omega$} & \multicolumn{1}{c}{$k$} \\
\hline
$^3$P$_1 \ - \ ^3$P$_2$ & M1 & 16831 &   3.2 \\
$^3$P$_0 \ - \ ^3$P$_2$ & E2 &  7514 &   0.5 \\
$^3$P$_1 \ - \ ^3$P$_0$ & M1 &  9317 &   5.3 \\
\end{tabular}
\end{ruledtabular}
\end{table}

Another interesting example is the positive ion of cerium (Ce~II).
An extract from the tables~\cite{martin} presenting experimental
energies and $g$-factors Lande of lower states of Ce~II are presented in 
Table~\ref{tabce} together with non-relativistic (NR) values of the 
$g$-factors. Non-relativistic $g$-factors were calculated according
to the formula
\begin{equation}
  g = 1 + \frac{J(J+1)-L(L+1)+S(S+1)}{2J(J+1)}.
\label{gfactor}
\end{equation}

The data presented in Table~\ref{tabce} reveals an interesting picture.
Fine structure multiplets of Ce~II intersect. There are many states with 
the same total momentum $J$ within an energy interval spanned by singe
fine structure multiplet. For example, there are four(!) states of $J=4.5$
within energy interval of the lowest fine structure multiplet $^4$H$^o$.
It is clear that they must be strongly mixed. Fine structure intervals
do not obey
Lande's rule and experimental $g$-factors deviate significantly
from the non-relativistic values. All these suggest strong configuration
mixing and sensitivity of the intervals to variation of $\alpha$.
An interval of particular interest is the $^4$H$^o_{11/2} \ - \ ^4$H$^o_{9/2}$
one. It is small, only 299~cm$^{-1}$ and states involved are strongly
mixed with other close states. This ensures strong enhancement of the 
change of $\alpha$ in the 
relative
change of frequency.

\begin{table}
\caption{\label{tabce}Energies (cm$^{-1}$) and $g$-factors 
of lower states of Ce~II.}
\begin{ruledtabular}
\begin{tabular}{cccrrr}
Config. & Term & $J$ & \multicolumn{1}{c}{$E$(exp)} & 
\multicolumn{1}{c}{$g$(exp)} & \multicolumn{1}{c}{$g$(NR)} \\
\hline
$4f5d^2$ & $^4$H$^o$ & 3.5 &      0.000 & 0.794 & 0.667 \\
         &           & 4.5 &   2581.257 & 1.023 & 0.970 \\
         &           & 5.5 &   2879.695 & 1.123 & 1.133 \\
         &           & 6.5 &   4203.934 & 1.189 & 1.231 \\
         &           &     &            &       &       \\
$4f5d^2$ &           & 4.5 &    987.611 & 0.948 & 1.000 \\
         &           &     &            &       &       \\
$4f5d^2$ & $^4$I$^o$ & 4.5 &   1410.304 & 0.856 & 0.727 \\
         &           & 5.5 &   2563.233 & 0.968 & 0.965 \\
         &           & 6.5 &   3793.634 & 1.128 & 1.107 \\
         &           & 7.5 &   5455.845 & 1.196 & 1.200 \\
         &           &     &            &       &       \\
$4f5d^2$ &           & 3.5 &   1873.934 & 0.806 & 1.000 \\
         &           &     &            &       &       \\
$4f5d^2$ &           & 0.5 &   2140.492 & 0.985 & 1.000 \\
         &           &     &            &       &       \\
$4f5d6s$ &           & 4.5 &   2382.246 & 1.039 & 1.000 \\
\end{tabular}
\end{ruledtabular}
\end{table}

Calculations for Ce~II are more difficult than for Te~I and Po~I  because
of valence states of high angular momentum ($5d$ and $4f$). 
Therefore we believe that it is premature to attempt them now.
The presence of enhancement is obvious but its actual value would
become important only on the stage of planning the measurements.
We are ready to perform the calculations if there is any interest 
from experimentalists.

Cases of fine structure anomalies similar to what is presented here for 
Ce~II can be easily fond in spectra of many other rare-earth elements. 
Which of them are suitable for the search of $\alpha$ variation is the
question which needs further consideration.

\section{Conclusion}

We present an alternative way to search for variation of the fine structure
constant in laboratory measurements. We propose to use fine structure 
intervals in the ground or low excited states of many-electron atoms which
are strongly perturbed by configuration interaction with neighboring states.
This method has double advantages. On one
 hand,
 the use of low states
ensures that they are metastable. This is important for very accurate
frequency measurements. On the other
hand, strong perturbation may lead to anomalously small fine structure
interval and
strong enhancement of the relative
 sensitivity of the frequencies to the change
of the fine structure constant.
Because of the high relative sensitivity one does need extremely accurate
absolute measurements of the frequencies (this is  the difference
with conventional atomic clock measurements).
Large value of the effect/frequency ratio  may also help to
 reduce importance of some systematic effects (e.g. the Doppler shift
 and broadening).   
Note, however, that we do not consider in this paper any practical
measurement scheme.

Enhanced and highly non-linear dependence of the small fine structure intervals
on the magnitude of the relativistic corrections
also presents certain theoretical interest.

\section{Acknowledgments}
    
    This work was supported by the Australian Research Council and Department
of Energy, Office of Nuclear Physics, Contract No. W-31-109-ENG-38.

\end{document}